\pdfoutput=1
\documentclass[pdflatex,iicol,sn-basic]{sn-jnl}

\usepackage[T1]{fontenc}
\usepackage[utf8]{inputenc}
\usepackage{lmodern}
\input{glyphtounicode}
\usepackage{microtype}%
\usepackage{amsmath,amssymb}
\usepackage{booktabs}
\usepackage{tabularx}
\newcolumntype{P}[1]{>{\raggedright\arraybackslash}p{#1}}%
\usepackage{graphicx}
\usepackage{xcolor}
\usepackage{algorithm}%
\usepackage{algpseudocode}
\usepackage{listings}%
\definecolor{aBlue}{HTML}{176B87}
\definecolor{aGreen}{HTML}{2E7D5B}
\definecolor{aGold}{HTML}{B8860B}
\definecolor{aGray}{HTML}{555555}
\definecolor{aRed}{HTML}{A03232}

\usepackage{orcidlink}%
\renewcommand{\orcid}[1]{\orcidlink{#1}}%
\usepackage{etoolbox}
\makeatletter
\patchcmd{\@maketitle}{\removelastskip\vskip36pt\vskip0pt}{\removelastskip\vskip12pt\vskip0pt}{\typeout{PATCH-1-OK}}{\typeout{PATCH-1-FAIL}}%
\patchcmd{\abstracthead}{-22pt \@plus0ex \@minus0ex}{-14pt \@plus0ex \@minus0ex}{\typeout{ABS-OK}}{\typeout{ABS-FAIL}}%
\patchcmd{\@maketitle}{\removelastskip\vskip24pt}{\removelastskip\vskip12pt}{\typeout{PATCH-2-OK}}{\typeout{PATCH-2-FAIL}}
\patchcmd{\@maketitle}{\removelastskip\vskip24pt}{\removelastskip\vskip12pt}{\typeout{PATCH-3-OK}}{\typeout{PATCH-3-FAIL}}
\patchcmd{\@maketitle}{\else\vskip21pt\fi}{\else\vskip9pt\fi}{\typeout{PATCH-4-OK}}{\typeout{PATCH-4-FAIL}}
\patchcmd{\@maketitle}{\global\punctcount\aucount\vskip20pt}{\global\punctcount\aucount\vskip8pt}{\typeout{PATCH-5-OK}}{\typeout{PATCH-5-FAIL}}
\renewcommand{\bibfont}{\small\@rightskip\@flushglue\rightskip\@rightskip}
\makeatother
\AtBeginDocument{\tolerance=2000\relax\setlength{\emergencystretch}{1.5em}}%
\makeatletter\AtBeginDocument{\@secpenalty=0\relax\setlength{\parskip}{0pt plus 1pt}}\makeatother%
\makeatletter\AtBeginDocument{\def\@textbottom{\vskip\z@\@plus 24pt}}\makeatother%
\usepackage{flushend}%
\AtBeginDocument{}%
\makeatletter\g@addto@macro\abstractfont{\mathversion{normal}}\makeatother%

\AtBeginDocument{%
  \hypersetup{%
    pdftitle={An open source Python code for three-dimensional topology optimization in single precision on a consumer GPU},
    pdfauthor={Leonilson Toffoli Pedrini; Rodolfo G. M. Andrade; Rafael M. Ferro},
    pdfkeywords={topology optimization, graphics processing units, stress constraints, robust formulation, compliant mechanisms, porous infill}%
  }%
}

\begin{document}
\articletype{Educational Article}%

\title[3D topology optimization in single precision on a consumer GPU]{An open source Python code for three-dimensional topology optimization in single precision on a consumer GPU}

\author*[1]{\fnm{Leonilson Toffoli} \sur{Pedrini}\orcid{0000-0002-9482-1007}}\email{leonilson.pedrini01@estudante.ifes.edu.br}
\author[1,2]{\fnm{Rodolfo G. M.} \sur{Andrade}\orcid{0000-0003-3956-5941}}\email{rodolfo.andrade@ifes.edu.br}
\author[1,3]{\fnm{Rafael M.} \sur{Ferro}\orcid{0000-0002-3229-2170}}\email{rafael.ferro@ifes.edu.br}

\affil*[1]{\orgdiv{Programa de P\'os-Gradua\c{c}\~ao em Engenharia Metal\'urgica e de Materiais (PROPEMM)}, \orgname{Instituto Federal do Esp\'irito Santo (IFES), Campus Vit\'oria}, \orgaddress{\street{Av. Vit\'oria 1729, Jucutuquara}, \city{Vit\'oria}, \state{ES}, \postcode{29040-780}, \country{Brazil}}}
\affil[2]{\orgdiv{Civil Engineering and Buildings Department}, \orgname{Instituto Federal do Esp\'irito Santo (IFES), Campus Vit\'oria}, \orgaddress{\street{Av. Vit\'oria 1729, Jucutuquara}, \city{Vit\'oria}, \state{ES}, \postcode{29040-780}, \country{Brazil}}}
\affil[3]{\orgdiv{Mechanical Engineering Department}, \orgname{Instituto Federal do Esp\'irito Santo (IFES), Campus Aracruz}, \orgaddress{\street{Av. Morob\'a 248, Morob\'a}, \city{Aracruz}, \state{ES}, \postcode{29192-733}, \country{Brazil}}}

\abstract{Three-dimensional topology optimization solves linear systems with millions of unknowns at every design
iteration. Consumer graphics cards run far faster in single than in double precision, and open codes for them mostly address minimum
compliance. This educational article presents TOF3D, an open source Python code for density-based topology
optimization in single precision on one consumer GPU. It solves four classes of problems: stress constrained minimum
compliance, robust minimum compliance against manufacturing errors, compliant mechanisms and bone-like porous infill.
The stiffness operator is applied without assembly, and the conjugate gradient method is preconditioned by geometric
multigrid. Rounding the element matrix to single precision gave rigid body motions a spurious energy and changed
compliances by up to 5.4\,\%. Forming the element products from differences of nodal displacements reduced this
error to ${1.3\times10^{-5}}$ at no cost in time. Adjoint sensitivities agreed with finite differences to within
${1.7\times10^{-6}}$, and the solutions agreed with Ansys Mechanical APDL to within ${4.1\times10^{-10}}$ with the same
element. On an RTX 3060 with 12\,GB, one design iteration of a cantilever took 0.46\,s with 0.5 million elements and
15\,s with 33.5 million. On the same card and problem, an open GPU code that applies its operator in double precision
took 3.7 and 21\,s with 0.5 and 4.2 million elements and ran out of memory on larger meshes. Eight cases fixed in a
protocol before the runs took 9 to 35 minutes each with 4.2 to 5.2 million elements.}


\maketitle

\section{Introduction}\label{sec:intro}

Topology optimization distributes material in a design domain to meet a structural goal under given loads and
supports \citep{bendsoe2003topology}. Besides stiffness, parts in service must
keep the stress below a limit and tolerate the erosion and dilation that manufacturing causes. Compliant mechanisms
must also deliver a prescribed motion, and parts built by additive manufacturing may carry a porous infill.
Density-based formulations exist for each of these requirements. The relaxed von Mises stress aggregated by a
$P$-norm handles the stress limit \citep{bruggi2008alternative,le2010stress,holmberg2013stress}, and the robust
formulation optimizes the eroded, intermediate and dilated designs together \citep{sigmund2009manufacturing,
wang2011projection}. For compliant mechanisms the objective is an output displacement \citep{sigmund1997compliant},
and for infill, local volume constraints produce a bone-like structure \citep{wu2018infill}. In three dimensions, however, each
design iteration solves one or more linear systems with millions of unknowns, and the cost of these solutions
limits the resolution at which the formulations are studied.

Educational codes made topology optimization accessible. The 99-line Matlab code of \citet{ferrari2020new} solves
two-dimensional minimum compliance problems in seconds, and its 125-line version extends them to three dimensions.
Open source implementations followed, built on general finite element libraries or on sparse solvers, some with an
optional GPU solver \citep{ferro2023opensource,kim2025pytopo3d}. For large meshes, a PETSc framework for computer
clusters produced giga-voxel designs \citep{aage2017giga}; at the other end, interactive apps brought two- and
three-dimensional optimization to handheld devices at coarse resolution
\citep{aage2013interactive,nobeljorgensen2015interactive3d}. \citet{wang2025matrixfree} combined a matrix-free operator with an adapted multigrid
preconditioner in Matlab and reached 128 million elements on a desktop computer with 64\,GB of memory, for minimum
compliance and porous infill.

Graphics processing units (GPUs) offer another route to large meshes. Several GPU codes applied the stiffness matrix element by element
or without assembling it \citep{schmidt2011gpu,martinezfrutos2015gpu,ratnakar2022ebe}, and \citet{wu2016system}
combined a matrix-free operator with a geometric multigrid preconditioner to reach high resolution on one GPU.
Clusters of GPUs extended these codes to larger meshes and to robust optimization \citep{martinezfrutos2016multigpu,
herreroperez2021multigpu}. \citet{traff2023gpu} released open codes in Futhark and OpenMP and optimized a
cantilever with 65.5 million elements over 100 design iterations in about two hours on one A100 GPU.
\citet{zhao2024fiber} wrote a multigrid code in Taichi for fiber-reinforced composites with 33.6 million elements,
and \citet{zhang2023homogenization} released a GPU solver for inverse homogenization. On a consumer card,
\citet{jia2026largescale} imposed length scale and local volume constraints with 151 million degrees of freedom. Most of
these works optimize stiffness.

Stress, robustness against manufacturing errors and mechanisms appear in fewer GPU works.
\citet{martinezfrutos2016multigpu} treated robustness as a probabilistic problem with random loads and material
properties, not as the erosion and dilation of the design. \citet{dugast2023support} optimized support structures
for metal additive manufacturing with the stress as objective on an A100 GPU, and verified their mechanical model against Ansys.
Most GPU codes also assume data center cards with large memory and fast double precision, whereas consumer cards
offer less memory and favor single precision \citep{traff2023gpu}. Finite element solvers on such hardware have
kept their accuracy by mixed precision iterative refinement, which computes the residual in double precision
\citep{goddeke2007mixed}. To the knowledge of the authors, no open code
combines stress constraints, robustness, compliant mechanisms and porous infill in three dimensions on a consumer GPU, with its responses cross-checked against a
commercial finite element code.

Thus, this work aims to provide TOF3D, an open source Python code developed by the authors for density-based
topology optimization in three dimensions, in single precision, on one consumer GPU. The code solves four classes of problems:
(i)~minimum compliance under a stress constraint, (ii)~robust minimum compliance against manufacturing errors,
(iii)~compliant mechanisms in standard and robust form, and (iv)~bone-like porous infill. The work makes
four contributions. The first is a single code, under the MIT license, that runs the four classes on a 12\,GB card. The
second is the verification: adjoint gradients checked against finite differences, and optimized designs
cross-checked against Ansys Mechanical APDL. The third is the time and memory of each class on meshes with millions
of elements, measured also against an open GPU code on the same card. These numbers serve as a reference for other GPU codes
and for learned surrogates, neural networks trained to predict optimized designs \citep{sigmund2022benchmarking,woldseth2022use}. The fourth concerns the single precision that consumer cards
favor. Rounding the element stiffness matrix to single precision gives rigid body motions a spurious energy, and on
meshes with millions of elements this changed compliances by up to 5.4\,\%. Computing the element products from
differences of nodal displacements reduces this error to ${1.3\times10^{-5}}$ at no cost in time and keeps the operator and all vectors in
single precision.

\section{Methods}\label{sec:methods}

TOF3D is an open source Python code, developed by the authors under the MIT license, that runs density-based
topology optimization of three-dimensional linear elastic structures on one GPU. Its six modules, about 2\,000 lines
in total, contain the finite element operator, the multigrid preconditioner, the density filter, the method of moving
asymptotes (MMA) and the four example problems. The GPU kernels are written in Taichi \citep{hu2019taichi}; NumPy and
SciPy serve only for input, output and the coarsest multigrid level. Each design iteration filters the design variables by fast Fourier transforms (FFT) and projects them, solves one to six linear systems by the conjugate gradient method with a multigrid preconditioner (MGCG), computes the sensitivities by the adjoint method, and updates the
design with the MMA (Algorithm~\ref{alg:iter}). Table~\ref{tab:comp} lists, for each component, the formulation it
follows, the module that implements it and its check.

\begin{table*}[!t]
\caption{Components of TOF3D: formulation followed, module written by the authors, and check. $\mathbf{K}^0$: element stiffness matrix for unit modulus; PCG: preconditioned conjugate gradient method. Third-party
software: Taichi 1.7.4 for the GPU kernels; NumPy 2.3 and SciPy 1.16 for input, output and the coarsest multigrid
level}\label{tab:comp}
\small\begin{tabular}{@{}P{0.24\textwidth}P{0.26\textwidth}P{0.16\textwidth}P{0.24\textwidth}@{}}
\toprule
Component & Formulation & Module & Check \\
\midrule
Eight-node hexahedron and matrix-free operator, Eqs.~(\ref{eq:simp})--(\ref{eq:diff}) & Trilinear element;
\citet{wu2016system} & \texttt{tof3d\_core} & Symmetry and rigid modes of $\mathbf{K}^0$ within $10^{-16}$; single
against double precision (Sect.~\ref{sec:rverif}) \\
Density filter and projection, Eqs.~(\ref{eq:filter})--(\ref{eq:proj}) & \citet{bourdin2001filters};
\citet{wang2011projection} & \texttt{tof3d\_filters}, \texttt{tof3d\_core} & FFT against direct convolution,
${4\times10^{-15}}$ \\
Problems and adjoint sensitivities, Eqs.~(\ref{eq:problem})--(\ref{eq:adjoint}) & \citet{sigmund1997compliant};
\citet{bruggi2008alternative}; \citet{le2010stress}; \citet{wu2018infill} & \texttt{tof3d\_core} & Central finite
differences (Sect.~\ref{sec:rverif}) \\
Geometric multigrid preconditioner, Eq.~(\ref{eq:galerkin}) & \citet{amir2014multigrid}; \citet{traff2023gpu} &
\texttt{tof3d\_mg} & Galerkin product ${4\times10^{-16}}$; MGCG against block Jacobi PCG ${1.6\times10^{-10}}$ \\
MMA with the dual subproblem on the GPU & \citet{svanberg1987mma}; \citet{aage2013parallel} & \texttt{tof3d\_mma},
\texttt{tof3d\_mma\_gpu} & Primal--dual interior point solver and SciPy, ${6\times10^{-7}}$ \\
Example problems & This work & \texttt{tof3d\_examples} & Protocol of Sect.~\ref{sec:pex} \\
\botrule
\end{tabular}
\end{table*}

\begin{algorithm}[!t]
\caption{One design iteration of TOF3D}\label{alg:iter}
\begin{algorithmic}[1]
\small
\State Filter $\boldsymbol\rho$ by FFT, Eq.~(\ref{eq:filter}), and project it for each design, Eq.~(\ref{eq:proj})
\State Set $E_e$ by Eq.~(\ref{eq:simp}) and rebuild the coarse operators, Eq.~(\ref{eq:galerkin})
\For{each load case}
\State Solve Eq.~(\ref{eq:equil}) by MGCG from the previous $\mathbf{u}$, with the operator of Eq.~(\ref{eq:diff})
\EndFor
\State Evaluate the objective and constraints of Eq.~(\ref{eq:problem}); solve the adjoint systems by MGCG
\State Form the sensitivities, Eq.~(\ref{eq:adjoint}), and carry them back through the projection and the filter
\State Update $\boldsymbol\rho$ by the MMA, with its dual subproblem solved on the GPU
\State Double $\beta$ at fixed iterations; robust problems refresh the volume bound $V_{\mathrm{d}}^{*}$ of the dilated design every 20 iterations
\end{algorithmic}
\end{algorithm}

\subsection{Computing environment}\label{sec:env}

All optimizations ran on one NVIDIA GeForce RTX 3060 with 12\,GB of memory (driver 595.91). The host was a desktop computer with an Intel Core i5-13500, 31\,GiB of memory and Ubuntu 24.04, running Python 3.12, Taichi 1.7.4 (CUDA backend), NumPy 2.3 and SciPy 1.16. The commercial cross-check used Ansys Mechanical APDL (MAPDL) 2025 R1 \citep{ansys2025solid185} on a second
computer with an Intel Core i9-13900K, 128\,GB of memory and Windows 11, with four cores per run.

Large language model assistants, including task-specific agents (Claude Code, Anthropic), were used to write and
review software, test and analysis scripts, to run the verification and timing studies, and to draft and restructure
the text. No reported quantity originates from a language model. Every value comes from the cited software and an
archived record. The authors reviewed the code, records, analyses and text and are responsible for the
interpretation and all claims.

\subsection{Finite element model and linear solver}\label{sec:fe}

The design domain is a regular grid of eight-node hexahedra of unit size with full ${2\times2\times2}$ integration and
Poisson ratio $\nu = 0.3$. The Young's modulus of element $e$ follows the solid isotropic material with penalization (SIMP) interpolation with a small minimum modulus
\citep{bendsoe1999material}
\begin{equation}
E_e = E_{\min} + \bar\rho_e^{\,3}\,(E_0 - E_{\min}), \label{eq:simp}
\end{equation}
with $E_0 = 1$, $E_{\min} = 10^{-4}$ and the physical density $\bar\rho_e$ of Sect.~\ref{sec:proj}. The stiffness matrix of the equilibrium system
\begin{equation}
\mathbf{K}(\bar{\boldsymbol\rho})\,\mathbf{u} = \mathbf{f} \label{eq:equil}
\end{equation}
is never assembled. Each GPU thread is assigned one node $i$ and gathers
\begin{equation}
(\mathbf{K}\mathbf{u})_i = \sum_{e \ni i} E_e \sum_{j \in e} \mathbf{K}^0_{ij}\,\mathbf{u}_j \label{eq:matfree}
\end{equation}
from its eight neighboring elements, where $\mathbf{K}^0_{ij}$ is the ${3\times3}$ block that couples nodes $i$ and
$j$ in the element stiffness matrix for unit modulus. This matrix-free scheme follows \citet{wu2016system} and
\citet{traff2023gpu}. Supports fix degrees of freedom, springs add to the diagonal, and elements outside the domain,
such as a cutout, carry $E_e = 0$. Vectors and $\mathbf{K}^0$ are stored in single precision,
and inner products are accumulated in double precision.

Rounding to single precision breaks one property of the element matrix. In exact arithmetic the blocks of each row of $\mathbf{K}^0$ sum to zero,
$\sum_{j}\mathbf{K}^0_{ij} = \mathbf{0}$, so a rigid translation of an element produces no nodal force. After rounding
to single precision the row sums reach about ${4\times10^{-9}}$, and rigid motions acquire a spurious stiffness. On a
uniform design the resulting error in compliance rose from ${2.3\times10^{-4}}$ on ${32\times16\times16}$ elements to ${4.9\times10^{-4}}$ on ${48\times24\times24}$, about $n^{1.9}$ between these two meshes with $n$ elements per side,
because displacements grow with the mesh while element strains do not. TOF3D therefore evaluates
Eq.~(\ref{eq:matfree}) in the difference form
\begin{equation}
(\mathbf{K}\mathbf{u})_i = \sum_{e \ni i} E_e \sum_{j \in e} \mathbf{K}^0_{ij}\,(\mathbf{u}_j - \mathbf{u}_i),
\label{eq:diff}
\end{equation}
which equals Eq.~(\ref{eq:matfree}) in exact arithmetic and vanishes for rigid translations whatever the rounding.
The multigrid smoother uses the same form, and the sensitivities and element stresses use displacements relative to
one node of the element. Unlike mixed precision refinement \citep{goddeke2007mixed}, Eq.~(\ref{eq:diff}) keeps the
operator and all vectors in single precision. The identity is elementary, and lattice Boltzmann codes use a similar step, storing each particle distribution relative to its lattice weight to limit round-off \citep{lehmann2022accuracy}. The authors found no earlier description of this form for matrix-free finite element operators in single precision. The present work reports its measured effect (Sect.~\ref{sec:rverif}).

The preconditioner of MGCG is one geometric multigrid V-cycle, following
\citet{amir2014multigrid} and \citet{traff2023gpu}. Each coarse level halves the elements in every direction. Its
operator is the Galerkin product
\begin{equation}
\mathbf{K}_{l+1} = \mathbf{P}_l^{\mathsf{T}}\,\mathbf{K}_l\,\mathbf{P}_l \label{eq:galerkin}
\end{equation}
with trilinear prolongation $\mathbf{P}_l$, built exactly by applying the finer operator to 81 probe vectors, three displacement components for each of 27 groups of coarse nodes whose stencils do not overlap. It is stored as a 27-point stencil of ${3\times3}$ blocks. Two sweeps of damped nodal block Jacobi smoothing ($\omega = 0.6$) precede and follow the coarse correction. The coarsest level, with at most 4\,500 degrees of freedom, is solved by a dense Cholesky factorization on the CPU. Near removed regions, the coarse operators come from an auxiliary operator in which the removed elements keep the modulus $E_{\min}$, so that the domain boundary cuts no coarse cell. The conjugate gradient method still solves the true system. This change cut the iterations per solution of the L-bracket example on ${256\times256\times80}$ elements (Sect.~\ref{sec:pex}) from 214 to 26. Each solution starts from the previous one of the same load case and stops at a relative
residual of $10^{-5}$; if MGCG has not converged after 500 iterations, the conjugate gradient method with a block Jacobi preconditioner (block Jacobi PCG) finishes the solution.

\subsection{Filter, projection and optimizer}\label{sec:proj}

A cone density filter of radius $r$ \citep{bourdin2001filters},
\begin{equation}
\begin{aligned}
\tilde\rho_e &= \frac{\sum_k w_{ek}\,\rho_k}{\sum_k w_{ek}},\\
w_{ek} &= \max\bigl(0,\, r - \lVert\mathbf{x}_e - \mathbf{x}_k\rVert\bigr),
\end{aligned}\label{eq:filter}
\end{equation}
with the sums over the elements of the domain, is evaluated by fast Fourier transforms. It is followed by the
smoothed Heaviside projection \citep{wang2011projection}
\begin{equation}
\bar\rho_e = \frac{\tanh(\beta\eta) + \tanh\bigl(\beta(\tilde\rho_e - \eta)\bigr)}
{\tanh(\beta\eta) + \tanh\bigl(\beta(1 - \eta)\bigr)} \label{eq:proj}
\end{equation}
with threshold $\eta$ and steepness $\beta$. The
robust formulation uses three projections: eroded, intermediate and dilated designs with thresholds
$\eta_{\mathrm{e}} > \eta_{\mathrm{i}} = 0.5 > \eta_{\mathrm{d}}$. The steepness doubles at fixed design iterations
(Table~\ref{tab:cases}), and the moving asymptotes of the MMA return to their initial positions at each doubling.

\begin{table*}[!t]
\caption{Cases of the protocol on the coarse mesh. $r$: filter radius in elements; $\beta$: final steepness, doubled
every 50 design iterations (60 for the two infill cases); move: MMA move limit}\label{tab:cases}
\small\begin{tabular}{@{}P{0.2\textwidth}P{0.2\textwidth}P{0.37\textwidth}ccc@{}}
\toprule
Case & Domain & Problem & $r$ & $\beta$ & Move \\
\midrule
L-bracket, compliance & ${96\times96\times32}$, cutout ${0.6L\times0.6L}$ & $V^{*} = 0.30$ & 2.5 & 8 & 0.1 \\
L-bracket, stress & same & $V^{*} = 0.30$, $\sigma_{\lim} = 0.70\,\sigma_{\max}$ of the compliance design & 2.5 & 8 & 0.1 \\
Cantilever, single & ${96\times48\times48}$ & $V^{*} = 0.12$, $\eta = 0.5$ & 3.0 & 32 & 0.2 \\
Cantilever, robust & same & $\eta = 0.75/0.5/0.25$ & 3.0 & 32 & 0.2 \\
Inverter, single & half domain ${96\times48\times48}$ & $V^{*} = 0.25$, springs $k_{\mathrm{in}} = k_{\mathrm{out}} = 1.6$ & 2.5 & 32 & 0.1 \\
Inverter, robust & same & $\eta = 0.7/0.5/0.3$ & 2.5 & 32 & 0.1 \\
Cantilever, infill & ${96\times48\times48}$ & $\alpha = 0.6$, $R = 6$, $p = 16$ & 1.5 & 16 & 0.2 \\
Cantilever, no infill & same & $V^{*}$ = final volume of the infill design & 1.5 & 16 & 0.2 \\
\botrule
\end{tabular}
\end{table*}

The design is updated by the MMA \citep{svanberg1987mma}, rewritten by the authors from the original equations. The
subproblem is solved through its dual in the $m$ constraint multipliers, with a Newton method whose elementwise
evaluations run on the GPU, as in the parallel MMA of \citet{aage2013parallel}. The approximations of the MMA carry a regularization term, $10^{-5}$ divided by the range of the design variables \citep{svanberg2007mma}, that keeps them strictly convex where a sensitivity vanishes. The force inverter
problems use the min-max form \citep{svanberg2007mma}, which minimizes an auxiliary variable that bounds the objective functions from above. When a design loses its load path between input and output, the barrier term of the min-max form dominates the dual Hessian and the remaining terms are lost to round-off. The Hessian is then singular, and the Newton step is the minimum norm least squares step; when the Hessian is regular, the update is unchanged.

\subsection{Four problems}\label{sec:problems}

\begin{table*}[!t]
\caption{The four problems. $\bar{\boldsymbol\rho}_{\mathrm{e}}$, $\bar{\boldsymbol\rho}_{\mathrm{i}}$,
$\bar{\boldsymbol\rho}_{\mathrm{d}}$: eroded, intermediate and dilated designs; $\hat\rho_e$: volume fraction in the
ball of radius $R$ around element $e$; $n$: number of elements; $V^{*}$, $\sigma_{\lim}$ and $\alpha$: bounds on the
volume fraction, the stress and the local volume fraction}\label{tab:problems}
\small\begin{tabular}{@{}P{0.15\textwidth}P{0.26\textwidth}P{0.34\textwidth}P{0.16\textwidth}@{}}
\toprule
Problem & Objective & Constraints & Reference \\
\midrule
Stress constrained & $C(\bar{\boldsymbol\rho}_{\mathrm{i}})$ & $V \le V^{*}$; $\bigl(\sum_e
(\sigma_e/\sigma_{\lim})^{P}\bigr)^{1/P} \le 1$ (normalized) & \citet{le2010stress} \\
Robust (cantilever) & $C(\bar{\boldsymbol\rho}_{\mathrm{e}})$ & $V(\bar{\boldsymbol\rho}_{\mathrm{d}}) \le
V_{\mathrm{d}}^{*}$ & \citet{wang2011projection} \\
Force inverter & $\max_{k\in\{\mathrm{e,i,d}\}} u_{\mathrm{out}}(\bar{\boldsymbol\rho}_k)$ (robust) or
$u_{\mathrm{out}}(\bar{\boldsymbol\rho}_{\mathrm{i}})$ & $V(\bar{\boldsymbol\rho}_{\mathrm{d}}) \le
V_{\mathrm{d}}^{*}$ or $V \le V^{*}$ & \citet{sigmund1997compliant,wang2011projection} \\
Porous infill & $C(\bar{\boldsymbol\rho}_{\mathrm{i}})$ & $\bigl(\frac{1}{n}\sum_e \hat\rho_e^{\,p}\bigr)^{1/p} \le
\alpha$ & \citet{wu2018infill} \\
\botrule
\end{tabular}
\end{table*}

All four problems take the form
\begin{equation}
\begin{aligned}
\min_{\boldsymbol\rho}\;\; & f_0(\bar{\boldsymbol\rho},\mathbf{u})\\
\text{s.t.}\;\; & \mathbf{K}(\bar{\boldsymbol\rho})\,\mathbf{u} = \mathbf{f},\\
& f_k(\bar{\boldsymbol\rho},\mathbf{u}) \le 0, \quad 0 \le \rho_e \le 1,
\end{aligned}\label{eq:problem}
\end{equation}
with the objective $f_0$ and the constraints $f_k$ of Table~\ref{tab:problems}. All minimize the compliance
$C = \mathbf{f}^{\mathsf{T}}\mathbf{u}$, except the force inverter, which minimizes the output displacement
$u_{\mathrm{out}} = \mathbf{l}^{\mathsf{T}}\mathbf{u}$, with $\mathbf{l}$ the vector that averages the displacement in the output direction over the output region, of a mechanism with input and output springs that represent the actuator and the workpiece
\citep{sigmund1997compliant}. The stress constraint bounds the $P$-norm
\begin{equation}
\begin{gathered}
\sigma_{\mathrm{PN}} = \Bigl(\sum_{e} \bigl(\sigma_e/\sigma_{\lim}\bigr)^{P}\Bigr)^{1/P},\\
\sigma_e = \bar\rho_e^{\,q}\,\sigma_{\mathrm{vM},e},
\end{gathered}
\label{eq:stress}
\end{equation}
of the relaxed von Mises stress, with the stress limit $\sigma_{\lim}$, $P = 10$, $q = 0.5$ and the adaptive
normalization of \citet{le2010stress}: the constraint is $c\,\sigma_{\mathrm{PN}} \le 1$, where $c$ is the mean of its previous value and $\max_e(\sigma_e/\sigma_{\lim})/\sigma_{\mathrm{PN}}$;
the relaxation follows \citet{bruggi2008alternative}. Elements within three element sizes of loads and supports are
excluded from the stress measure. The robust problems minimize the worst of the eroded, intermediate and dilated
designs,
\begin{equation}
\min_{\boldsymbol\rho}\; \max_{k\in\{\mathrm{e},\mathrm{i},\mathrm{d}\}} f_0(\bar{\boldsymbol\rho}_k),
\label{eq:robust}
\end{equation}
and constrain the volume of the dilated design. Its bound $V_{\mathrm{d}}^{*}$ is refreshed every 20 design
iterations so that the intermediate design meets $V^{*}$ \citep{wang2011projection}. The infill problem bounds the
$p$-mean of the local volume fractions \citep{wu2018infill},
\begin{equation}
\Bigl(\frac{1}{n}\sum_e \hat\rho_e^{\,p}\Bigr)^{1/p} \le \alpha, \qquad
\hat\rho_e = \frac{1}{|\mathcal{N}_e|}\sum_{k\in\mathcal{N}_e}\bar\rho_k, \label{eq:infill}
\end{equation}
with the bound $\alpha$, $p = 16$ and $\mathcal{N}_e$ the elements within a ball of radius $R$ around element $e$. The sensitivities
follow from the adjoint method. For a function $f(\bar{\boldsymbol\rho},\mathbf{u})$,
\begin{equation}
\frac{\mathrm{d}f}{\mathrm{d}\bar\rho_e} = \frac{\partial f}{\partial\bar\rho_e}
- \boldsymbol\lambda^{\mathsf{T}}\frac{\partial\mathbf{K}}{\partial\bar\rho_e}\,\mathbf{u}, \qquad
\mathbf{K}\boldsymbol\lambda = \frac{\partial f}{\partial\mathbf{u}}, \label{eq:adjoint}
\end{equation}
and the chain rule carries them back through Eqs.~(\ref{eq:proj}) and (\ref{eq:filter}). The adjoint systems use
the same MGCG.

The objective functions of the force inverter are divided by one tenth of the initial $|u_{\mathrm{out}}|$, because with the physical scale alone the element sensitivities ($\approx 10^{-7}$) fell below the regularization of the MMA and no mechanism formed. They are then shifted by 500, because the min-max form of the MMA requires positive functions; the shift does not change the sensitivities.

Each problem is defined by a short NumPy function that marks supports, loads, springs and passive elements on the
node and element grids. Listing~\ref{lst:cantilever} shows the cantilever. A new problem needs such a function, a
branch in the problem setup function \texttt{build} and an entry in the table of cases of \texttt{tof3d\_examples.py}.

\begin{lstlisting}[float=*t, caption={Definition of the cantilever in \texttt{tof3d\_examples.py}. \texttt{free}:
supports (0 = fixed degree of freedom); \texttt{f} and \texttt{f\_lateral}: nominal and lateral load cases;
\texttt{solid}: passive solid elements under the load; \texttt{s}: mesh ratio $N/96$; \texttt{C}: module \texttt{tof3d\_core}; \texttt{fe}: finite element grid; \texttt{In}, \texttt{Jn}, \texttt{Kn}: node indices; \texttt{ld}: loaded nodes; \texttt{act}: active elements; \texttt{smask}: elements of the stress measure, unused here; \texttt{n\_carga}: number of loaded nodes}, label=lst:cantilever]
def geo_cantilever(fe, s):
    In, Jn, Kn = C.node_grid(fe)
    free = np.ones(fe.nshape + (3,))
    free[0] = 0.0
    zc = fe.nz // 2
    w = int(round(2 * s))
    ld = (In == fe.nx) & (Jn <= w) & (np.abs(Kn - zc) <= w)
    f = np.zeros(fe.nshape + (3,))
    f[ld, 1] = -1.0 / ld.sum()
    flat = np.zeros(fe.nshape + (3,))
    flat[ld, 2] = 1.0 / ld.sum()
    solid = C.elems_touching(fe, ld)
    act = np.ones(fe.eshape, bool)
    return dict(act=act, free=free, spring=None, loads={"f": f, "f_lateral": flat}, solid=solid,
                void=np.zeros(fe.eshape, bool), smask=None, info=dict(n_carga=int(ld.sum())))
\end{lstlisting}

\subsection{Verification protocol}\label{sec:pverif}

The adjoint sensitivities of the five problem functions were compared with central finite differences (step $10^{-6}$) in double precision, on a ${6\times4\times3}$ grid with fixed and spring supports. The criterion was a relative difference of at most $10^{-5}$. The compliance and output displacement were checked again with MGCG on an ${8\times4\times4}$ grid. Three further comparisons checked the building blocks: block Jacobi PCG against a direct solution, the Galerkin coarse operators against explicit products $\mathbf{P}^{\mathsf{T}}\mathbf{K}\mathbf{P}$, and the FFT filter against direct convolution. The GPU dual solver of the MMA was compared with a primal--dual interior point solver and with SciPy.

The effect of single precision was measured on the 16 final designs of the examples. Each design was evaluated three
times: in single precision at a relative residual of $10^{-5}$, as in production, and in double precision at
$10^{-5}$ and at $10^{-12}$. The criterion, fixed before the evaluation, was a relative difference between single
precision and double precision at $10^{-12}$ of at most $10^{-3}$ in compliances and output displacements and
$10^{-2}$ in peak stress. The eight cases of the coarse mesh were also optimized entirely in double precision, for
information, without a criterion.

\subsection{Example protocol}\label{sec:pex}

Eight cases, a reference and a formulation under study for each problem, were fixed in a protocol before the runs, together
with three execution criteria and four hypotheses (Table~\ref{tab:cases}). Each case ran for 300 design iterations on
a coarse mesh (L-bracket ${96\times96\times32}$, other cases ${96\times48\times48}$ elements) and a fine mesh
(${256\times256\times80}$ and ${256\times128\times128}$), with every length of the problem scaled by the mesh ratio. The total load stays at one, and the spring stiffness of the inverter grows with the width of its patch: 1.6, 4.4 and 8.6 on the meshes of 96, 256 and 512 elements in length. The three execution criteria were 300 design iterations without invalid values, a constraint violation of at most 0.01, and a mean grayness, the mean of $4\bar\rho_e(1-\bar\rho_e)$ over the design, of at most 0.20 for the L-bracket and 0.10 otherwise. The four hypotheses were as follows. H1: the stress constrained L-bracket lowers the peak stress to at most 0.75 of
the compliance design, with at most 15\,\% more compliance. H2: the eroded robust cantilever has lower compliance than
the eroded single projection cantilever, and a morphological opening, an erosion followed by a dilation with a ball that deletes members thinner than the ball, removes less solid from it. H3: under erosion
($\eta = 0.70$), the single projection inverter loses more than half of its output displacement, while the robust one
loses less than half and keeps the output sign in all three designs. H4: the infill cantilever has lower lateral
compliance than a solid design of the same volume, with at most 1.5 times its nominal compliance. Verdicts were recorded as obtained. Changes to the protocol were recorded as dated amendments with their reasons, listed in Online Resource~1 and in
the deposit. A first run of the protocol used the code without the difference form of Eq.~(\ref{eq:diff}). All examples reported here come from a second run with the corrected code, which has that form; the first run verdicts are kept in the deposit. The second run was therefore not blind to the verdicts of the first, although the hypotheses and their limits did not change. A secondary analysis repeated the two L-bracket cases on the coarse mesh with three variants of the linear solver: tolerances of $10^{-6}$ and $10^{-7}$, and block Jacobi PCG. The optimization in double precision counts as a fourth variant. By the rule fixed for this analysis, a clause of H1 that holds in some of the runs but not in all is decided by the optimization path, and its range is reported. The single projection cantilever also ran on ${512\times256\times256}$ elements (Amendment 2). After the protocol, five more cases ran on this mesh as additional examples, without hypotheses: the robust cantilever, the two infill cases and the two inverters. Amendment 10 added two analyses after these runs: an opening with a radius scaled by the mesh ratio, and a test of the regularization of the MMA on the coarse inverter. Amendment 11 repeated the two inverters on this mesh with that regularization scaled, under criteria fixed before the runs.

\subsection{Commercial cross-check protocol}\label{sec:pansys}

The cross-check used the final intermediate designs of the first run of the protocol: the eight coarse cases and the two cantilevers on the fine mesh. They were thresholded at $\rho = 0.5$, with $E = 1$ in the solid and $E_{\min}$ in the void. Each design was written as an MAPDL input file with the same nodes, supports, loads and springs. MAPDL uses SOLID185 with B-bar integration, a selective reduced integration of the volumetric strain \citep{ansys2025solid185}, whereas the
production element of TOF3D is fully integrated. TOF3D therefore solved each coarse case twice, once with each element, in double precision to a relative residual of $10^{-12}$. The run with the production element quantifies the effect of the element formulation; the run with the B-bar element matrix of SOLID185 compares the two codes with the same element. The fine cases used the B-bar element only. The B-bar variant needs $\omega = 0.4$
in the smoother, because the largest eigenvalue of $\mathbf{D}^{-1}\mathbf{K}$, with $\mathbf{D}$ the block diagonal of $\mathbf{K}$, reaches $\lambda_{\max} = 3.9$ and the
damped smoother converges only for $\omega\lambda_{\max} < 2$. MAPDL used the sparse direct solver on the coarse mesh
and the PCG solver with a tolerance of $10^{-10}$ on the fine mesh. The limits on the relative differences, fixed before the runs, were $10^{-8}$ in compliance, $10^{-7}$ in nodal
displacements and in the output displacement of the inverter, and $10^{-6}$ in element stresses on the coarse mesh,
and $10^{-6}$, $10^{-5}$ and $10^{-4}$ on the fine mesh.

\begin{table*}[!t]
\caption{Worst relative difference between adjoint sensitivities and central finite differences, in double
precision}\label{tab:verif}
\small\begin{tabular}{@{}llc@{}}
\toprule
Function & Linear solver & Worst relative difference \\
\midrule
Compliance & Block Jacobi PCG & ${5.6\times10^{-8}}$ \\
Compliance of the eroded design (three projections) & Block Jacobi PCG & ${1.5\times10^{-8}}$ \\
Output displacement of the force inverter & Block Jacobi PCG & ${3.3\times10^{-7}}$ \\
$P$-norm of the relaxed von Mises stress & Block Jacobi PCG & ${8.8\times10^{-7}}$ \\
Local volume ($p$-mean) & Block Jacobi PCG & ${4.7\times10^{-9}}$ \\
Compliance & MGCG & ${8.6\times10^{-7}}$ \\
Output displacement of the force inverter & MGCG & ${1.7\times10^{-6}}$ \\
\botrule
\end{tabular}
\end{table*}

\subsection{Timing protocol}\label{sec:ptime}

The time per design iteration is the median over the 300 design iterations; the GPU memory is
the peak of samples taken every 30\,s. The linear solver was timed separately on the cantilever, with the optimized single projection
design of the fine mesh from the first run carried to seven meshes from ${64\times32\times32}$ to ${512\times256\times256}$ elements. Each time is the median of five solutions from a zero start to a relative residual of $10^{-5}$. Block Jacobi PCG was timed up to ${384\times192\times192}$ elements.

The open Futhark code of \citet{traff2023gpu} (version 0.9.1, compiled with Futhark 0.21.11 for OpenCL) ran on the
same card with its default problem. This is a cantilever with a load on a half disc at the tip, a volume fraction of 0.2 and
a filter radius of 1.5 elements. Only its backend and the output frequency of its design files were changed. TOF3D
solved the same problem with the same element, filter, supports, loads, residual tolerance of $10^{-5}$ and stopping
rule. Only the optimizers differed: optimality criteria in their code and the MMA in TOF3D, both with a move limit
of 0.1. Four meshes went from ${128\times64\times64}$ to ${512\times256\times256}$ elements, with 20 design iterations
and three repetitions per code, alternating the codes after one warm-up run of each. Before any time was compared,
the compliance of the first design iteration of each code had to agree with a double precision TOF3D reference
(relative residual $10^{-10}$) to within ${2\times10^{-3}}$. The time per design iteration is the median of iterations 2 to 20.

\subsection{Limitations of the method}\label{sec:lim}

TOF3D works on structured grids of hexahedra. The analysis is linear elastic, including the mechanisms, so it does not model large displacements. The 12\,GB of GPU memory limit the problem size: the largest runs, on ${512\times256\times256}$ elements, used 9.4 to 9.9\,GiB, and the L-bracket was run only up to ${256\times256\times80}$ elements. The coarsest multigrid level runs on the CPU and dominates
the solution time on small meshes. In some design iterations the MMA dual solver reached its cap of 300 Newton steps,
and the inverter designs passed through states without a load path before forming a mechanism. The regularization of the MMA in the released code does not scale with the number of elements: on the ${512\times256\times256}$ mesh the force inverters formed a mechanism only with a value set for that mesh, and only one value was tried (Sect.~\ref{sec:rtime}).

\section{Results and discussion}\label{sec:results}
\subsection{Verification}\label{sec:rverif}

The adjoint sensitivities of the five problem functions agreed with central finite differences to within
${8.8\times10^{-7}}$, below the criterion of $10^{-5}$ (Table~\ref{tab:verif}). The $P$-norm stress gave the largest
error; compliance, the eroded compliance and the local volume agreed to within ${6\times10^{-8}}$. With the multigrid
preconditioner in place of block Jacobi, the sensitivities stayed within ${1.7\times10^{-6}}$. The Galerkin coarse
operators reproduced $\mathbf{P}^{\mathsf{T}}\mathbf{K}\mathbf{P}$ to round-off (${4\times10^{-16}}$), and the MGCG
and block Jacobi solutions of the same systems differed by at most ${1.6\times10^{-10}}$ at a relative residual of
$10^{-10}$. The GPU dual solver of the MMA subproblem matched the reference primal--dual interior point solver to within ${5\times10^{-8}}$ after one design iteration and ${6\times10^{-7}}$ after five. The FFT density filter matched direct convolution to within ${4\times10^{-15}}$.

\begin{table*}[!t]
\caption{Largest relative difference between single precision (relative residual $10^{-5}$) and double precision
($10^{-12}$) over the final designs, before and after the difference form of Eq.~(\ref{eq:diff}). Before: first run designs;
after: the designs reported here}\label{tab:precision}
\small\begin{tabular*}{\textwidth}{@{\extracolsep{\fill}}lllll@{}}
\toprule
 & \multicolumn{2}{c}{Before} & \multicolumn{2}{c}{After} \\
\cmidrule(lr){2-3}\cmidrule(lr){4-5}
Response & Coarse & Fine & Coarse & Fine \\
\midrule
Compliance & ${3.7\times10^{-3}}$ & ${1.9\times10^{-2}}$ & ${9.1\times10^{-7}}$ & ${4.8\times10^{-7}}$ \\
Lateral compliance & ${7.9\times10^{-3}}$ & ${5.4\times10^{-2}}$ & ${1.1\times10^{-6}}$ & ${1.5\times10^{-6}}$ \\
Peak stress & ${3.3\times10^{-3}}$ & ${2.8\times10^{-2}}$ & ${4.5\times10^{-6}}$ & ${1.2\times10^{-5}}$ \\
Output displacement & ${7.6\times10^{-4}}$ & ${5.9\times10^{-3}}$ & ${8.1\times10^{-7}}$ & ${1.3\times10^{-5}}$ \\
\botrule
\end{tabular*}
\end{table*}

With the difference form of Eq.~(\ref{eq:diff}), single precision changed the responses of the 16 final designs by at most
${1.3\times10^{-5}}$, far below the criteria (Table~\ref{tab:precision}). Without it, the same evaluation of the
first run designs on the fine mesh gave errors of up to 5.4\,\% in the lateral compliance of the robust cantilever.
The peak stress of the stress constrained L-bracket differed by 2.8\,\%. The errors grew five- to ninefold from the coarse
to the fine mesh. The corrected code evaluated the same first run designs to within ${8.6\times10^{-6}}$, and one
application of the operator took 1.98\,ms instead of 2.02\,ms on the ${256\times128\times128}$ mesh. Double precision
with a residual of $10^{-12}$ instead of $10^{-5}$ changed the responses by at most ${9.7\times10^{-6}}$. Optimized entirely in double precision, the eight coarse cases gave the same verdicts for H2 to H4, and their final responses differed from single precision by at most 0.54\,\%, except the compliance of the stress constrained L-bracket (Sect.~\ref{sec:rex}).

\subsection{Examples}\label{sec:rex}

All 16 optimizations, eight cases on each mesh, met the three execution criteria of the protocol. They completed 300 design iterations without invalid values, violated their constraints by at most ${5\times10^{-4}}$, and ended with a mean grayness of at most 0.067. On the fine mesh the four hypotheses
were confirmed. On the coarse mesh H2 and H4 were confirmed, H3 was not, and the compliance clause of H1 held in the protocol run
but not in three variants of the linear solver or in double precision, so its verdict depends on the optimization path (Table~\ref{tab:hyp}).

\begin{table*}[!t]
\caption{Hypotheses of the protocol on the coarse mesh (L-bracket ${96\times96\times32}$, other cases
${96\times48\times48}$ elements) and the fine mesh (${256\times256\times80}$ and ${256\times128\times128}$). Ratios are
taken between the formulation under study and its reference}\label{tab:hyp}
\small\begin{tabular}{@{}lllll@{}}
\toprule
Hypothesis & Quantity & Required & Coarse & Fine \\
\midrule
H1, stress & Peak stress ratio & $\le 0.75$ & 0.700 & 0.700 \\
 & Compliance ratio & $\le 1.15$ & 1.120 to 1.265$^{1}$ & 1.036 \\
 & Verdict & & path dependent$^{1}$ & confirmed \\
H2, robust & Eroded compliance ratio & $< 1$ & 0.947 & 0.692 \\
 & Solid removed by opening$^{2}$ & robust $<$ single & 8.6\,\% $<$ 27.9\,\% & 0.89\,\% $<$ 0.96\,\% \\
 & Verdict & & confirmed & confirmed \\
H3, mechanism & Output loss at $\eta = 0.70$, single & $> 50\,\%$ & 24\,\% & 98\,\% \\
 & Output loss at $\eta = 0.70$, robust & $< 50\,\%$ & 0.5\,\% & 1.7\,\% \\
 & Verdict & & not confirmed & confirmed \\
H4, infill & Lateral compliance ratio & $< 1$ & 0.957 & 0.922 \\
 & Nominal compliance ratio & $\le 1.5$ & 1.163 & 1.158 \\
 & Verdict & & confirmed & confirmed \\
\botrule
\end{tabular}

\vspace{2pt}{\footnotesize $^{1}$1.120 in the protocol run; three variants of the linear solver and an optimization in double precision gave 1.242 to 1.265, so the compliance clause
held in one of five runs; 1.152, not confirmed, in the first run of the protocol (Sect.~\ref{sec:pex}).
$^{2}$Opening of the solid ($\rho > 0.5$) with a ball of radius two elements}\par
\end{table*}
\begin{figure}[!tb]
\centering
\includegraphics[width=\columnwidth]{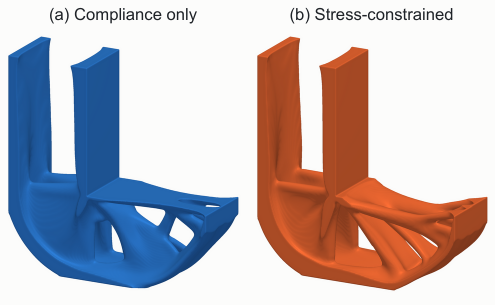}
\caption{L-bracket on the fine mesh: (a) compliance design; (b) stress constrained design. Isosurface $\rho = 0.5$
of the intermediate design}\label{fig:pairL}
\end{figure}
\begin{figure*}[!t]
\centering
\includegraphics[width=0.72\textwidth]{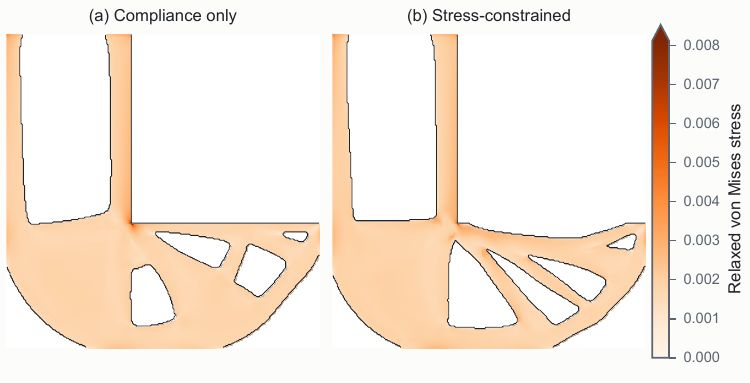}
\caption{Relaxed von Mises stress of the L-bracket on the fine mesh, maximum through the thickness, on the same
scale. The outline marks the solid ($\rho > 0.5$)}\label{fig:stress}
\end{figure*}

The stress constraint lowered the peak relaxed von Mises stress of the L-bracket to 0.700 of that of the compliance
design on both meshes, which is the value the constraint imposes. The increase in compliance differed between the
meshes: 3.6\,\% on the fine mesh and 12.0\,\% on the coarse mesh. On the coarse mesh, the three variants of the
linear solver and the optimization in double precision raised this increase to 24 to 26\,\%, above the 15\,\% that H1 allows, with the same stress ratio of 0.700.
The compliance designs of the five runs differed by less than 0.01\,\%, so the scatter comes from the
stress constrained design. The verdict of H1 on the coarse mesh therefore depends on the optimization path, as it did in
the first run of the protocol, where the ratio ranged from 1.124 to 1.152. The tolerance alone does not explain the scatter: tolerances of $10^{-6}$ and $10^{-7}$ gave 1.263 and 1.265, but block Jacobi and double precision, both at $10^{-5}$, gave 1.242 and 1.257.
The stress constrained design (Fig.~\ref{fig:pairL}) rounds the re-entrant corner, moving material away from it, and splits the horizontal
arm into more members. This spreads the stress that the compliance design concentrates at the corner
(Fig.~\ref{fig:stress}).
The same features appear in two dimensions: the stress constrained L-brackets of \citet{le2010stress} move
material away from the re-entrant corner and open the arm into a truss, and \citet{holmberg2013stress} obtained
L-beam designs that differ markedly from the compliance design and avoid the stress concentration at the corner,
with 6\,400 elements.

\begin{figure}[!tb]
\centering
\includegraphics[width=\columnwidth]{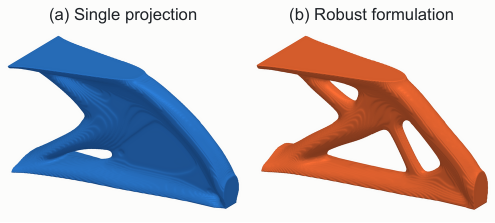}
\caption{Cantilever on the fine mesh: (a) single projection; (b) robust formulation. Isosurface $\rho = 0.5$ of the
intermediate design}\label{fig:pairC}
\end{figure}
\begin{figure}[!tb]
\centering
\includegraphics[width=\columnwidth]{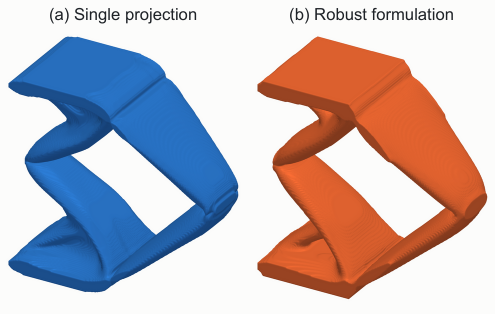}
\caption{Force inverter on the fine mesh: (a) single projection; (b) robust formulation. Isosurface $\rho = 0.5$ of
the intermediate design, mirrored about the symmetry plane}\label{fig:pairM}
\end{figure}
\begin{figure*}[!t]
\centering
\includegraphics[width=\textwidth]{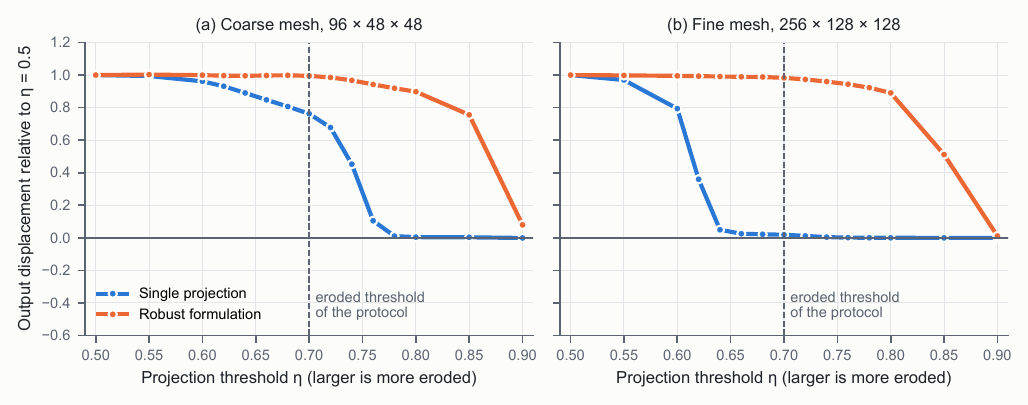}
\caption{Output displacement of the force inverter against the projection threshold $\eta$, relative to its value at
$\eta = 0.5$. Every linear solution reached the residual tolerance}\label{fig:erosion}
\end{figure*}

The robust formulation traded nominal stiffness for tolerance to erosion (Fig.~\ref{fig:pairC}). On the fine mesh the eroded robust
cantilever had 0.692 times the compliance of the eroded single projection cantilever, while its intermediate design had 6.3\,\% more compliance. On the coarse mesh these numbers were 0.947 and 9.0\,\%. A morphological opening with a
ball of radius two elements removed less solid from the robust design on both meshes (Table~\ref{tab:hyp}), in line
with the length scale control of the robust formulation \citep{wang2011projection,lazarov2016length}. A ball of fixed radius becomes small against the members as the mesh is refined. With the radius scaled by the mesh ratio, five elements on the fine mesh, the opening removed 7.8\,\% of the robust design and 17.7\,\% of the single projection design (Amendment 10).

The force inverter shows the same contrast more sharply (Fig.~\ref{fig:pairM}). On the fine mesh, eroding the
single projection inverter to $\eta = 0.70$ removed 98\,\% of its output displacement, while the robust inverter lost
1.7\,\%. On the coarse mesh the single projection inverter lost 24\,\% at $\eta = 0.70$ and more than half only at
$\eta = 0.74$, so H3, which requires a loss above one half at $\eta = 0.70$, was not confirmed. The robust inverter
kept at least 89\,\% of its output displacement up to $\eta = 0.80$ on both meshes.
Erosion cuts the thin necks that act as hinges in the single projection inverters,
the weakness of such mechanisms that the robust formulation was designed to remove
\citep{sigmund2009manufacturing,wang2011projection}. Figure~\ref{fig:erosion} gives the output displacement over the whole range of erosion.

\begin{figure}[!tb]
\centering
\includegraphics[width=\columnwidth]{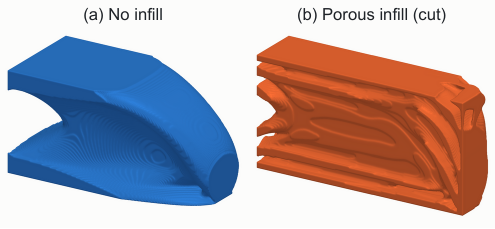}
\caption{Cantilever on the fine mesh, same volume: (a) without infill; (b) with porous infill, rear half of the
width. Isosurface $\rho = 0.5$ of the intermediate design}\label{fig:pairI}
\end{figure}

The porous infill reduced the lateral compliance of the cantilever to 0.922 of that of the reference with the same
volume, at 1.158 times its nominal compliance (0.957 and 1.163 on the coarse mesh). The local volume reached at most
0.87 in the infill design, against 1.0 in the reference, and the interior consists of thin walls parallel to the
long axis (Fig.~\ref{fig:pairI}; Online Resource~1, Fig.~S1). This is the trade-off between nominal stiffness and robustness to load
changes that \citet{wu2018infill} reported for bone-like infill.

\subsection{Cross-check against MAPDL}\label{sec:ransys}

\begin{table*}[!t]
\caption{Largest relative differences between TOF3D and MAPDL. Compliance $C$; nodal displacements $\mathbf{u}$
(Euclidean norm); element von Mises stress $\sigma_{\mathrm{vM}}$ (largest difference over the solid elements,
relative to the peak stress); output displacement $u_{\mathrm{out}}$ of the force inverter}\label{tab:mapdl}
\small\begin{tabular}{@{}lllllll@{}}
\toprule
TOF3D element & Mesh (cases) & MAPDL solver & $C$ & $\mathbf{u}$ & $\sigma_{\mathrm{vM}}$ & $u_{\mathrm{out}}$ \\
\midrule
B-bar & Coarse (8) & Sparse direct & ${2.4\times10^{-11}}$ & ${3.0\times10^{-10}}$ & ${2.3\times10^{-8}}$ & ${7.8\times10^{-12}}$ \\
B-bar & Fine (2) & PCG, $10^{-10}$ & ${3.8\times10^{-11}}$ & ${4.1\times10^{-10}}$ & ${1.6\times10^{-8}}$ & -- \\
Production & Coarse (8) & Sparse direct & 0.32--1.0\,\% & -- & 0.07--4.6\,\%$^{1}$ & 0.12--0.31\,\% \\
\botrule
\end{tabular}

\vspace{2pt}{\footnotesize $^{1}$Difference in peak stress, not element by element}\par
\end{table*}

With the same element on both sides, TOF3D and MAPDL agreed to within ${4\times10^{-11}}$ in compliance and
${4.1\times10^{-10}}$ in nodal displacements, in all ten cases and on both meshes (Table~\ref{tab:mapdl}). The element
stresses agreed to within ${2.3\times10^{-8}}$, below the floor of about ${6\times10^{-8}}$ set by the single precision
storage of element results in the MAPDL result file. The largest comparison had 12.8 million degrees of freedom,
where MAPDL used its PCG solver at a tolerance of $10^{-10}$. All differences stayed at least 40 times below the limits fixed before the runs
(Sect.~\ref{sec:pansys}). The cross-check runs in double precision and does not depend on the difference form of
Eq.~(\ref{eq:diff}): repeated with the corrected code, the eight coarse cases again agreed with MAPDL to within
${5\times10^{-12}}$ in compliance.

The production element of TOF3D, the fully integrated eight-node hexahedron, differs from SOLID185 with B-bar
integration by 0.32 to 1.0\,\% in compliance and by 0.07 to 4.6\,\% in peak stress. This difference comes from the
element formulation, since the two codes agree when they use the same element. \citet{dugast2023support} verified their mechanical model against Ansys;
the present cross-check covers ten optimized designs from the four problems and separates the error of the code, at most
${4.1\times10^{-10}}$ in displacement, from the effect of the element.

\subsection{Time and memory on a consumer GPU}\label{sec:rtime}

\begin{figure*}[tp]
\centering
\includegraphics[width=\textwidth]{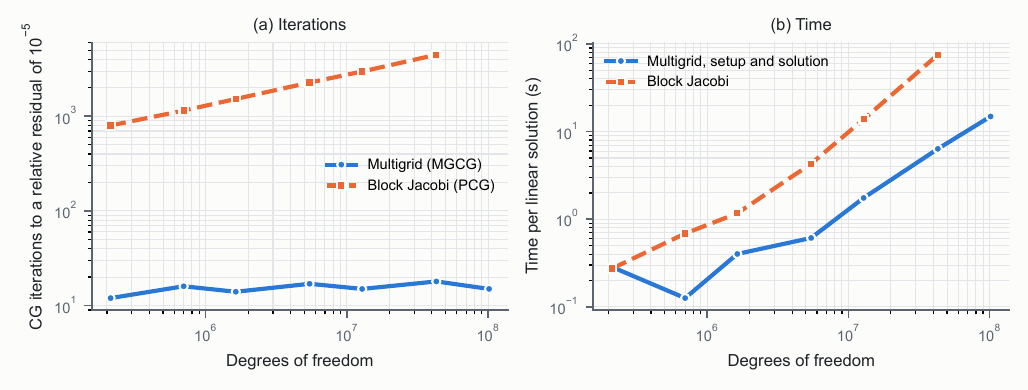}
\caption{Linear solver on the RTX 3060 for the cantilever, with the optimized design of the fine mesh carried to each
mesh. Median of five measurements}\label{fig:scaling}
\end{figure*}

The number of MGCG iterations stayed between 12 and 18 from 0.2 to 101.6 million degrees of freedom (Fig.~\ref{fig:scaling}). Block Jacobi PCG needed from 800 to 4\,475 iterations up to 43 million degrees of freedom. From
12.8 million degrees of freedom on, one linear solution with MGCG, setup included, was 8 to 12 times faster
than with block Jacobi.
On the smallest meshes the coarsest level dominates. Its dense Cholesky solution runs on the CPU at about 10\,ms per V-cycle with ${16\times8\times8}$ coarse elements, and at 1.6\,ms with ${12\times6\times6}$. The mesh with 0.7 million degrees of freedom has the smaller coarsest level and therefore solves faster than those with 0.2 and 1.6 million.

\begin{table}[!t]
\caption{Time per design iteration (s, median), total time (min) and peak GPU memory (GiB) of the optimizations on
the RTX 3060, 300 design iterations each. Fine mesh of Table~\ref{tab:cases} and the mesh of
${512\times256\times256}$ elements}\label{tab:time}
\footnotesize\setlength{\tabcolsep}{3.5pt}\begin{tabular}{@{}lrrrc@{}}
\toprule
Case & Elements & s/it. & min & GiB \\
\midrule
L-bracket, compliance & 5.2\,M & 2.5 & 12.9 & 1.61 \\
L-bracket, stress & 5.2\,M & 4.6 & 24.6 & 1.64 \\
Cantilever, single & 4.2\,M & 2.0 & 11.6 & 1.33 \\
Cantilever, robust & 4.2\,M & 2.1 & 10.8 & 1.39 \\
Inverter, single & 4.2\,M & 2.9 & 16.6 & 1.36 \\
Inverter, robust & 4.2\,M & 7.2 & 35.4 & 1.43 \\
Cantilever, infill & 4.2\,M & 2.4 & 13.1 & 1.39 \\
Cantilever, no infill & 4.2\,M & 1.7 & 8.9 & 1.39 \\
\midrule
Cantilever, single & 33.5\,M & 17.0 & 86.5 & 9.43 \\
Cantilever, robust & 33.5\,M & 19.0 & 98.1 & 9.43 \\
Inverter, single & 33.5\,M & 17.8 & 93.3 & 9.55 \\
Inverter, robust & 33.5\,M & 43.8 & 222.9 & 9.93 \\
Inverter, single$^{1}$ & 33.5\,M & 23.4 & 124.6 & 9.52 \\
Inverter, robust$^{1}$ & 33.5\,M & 60.6 & 308.6 & 9.90 \\
Cantilever, infill & 33.5\,M & 21.1 & 108.1 & 9.43 \\
Cantilever, no infill & 33.5\,M & 15.5 & 77.9 & 9.43 \\
\botrule
\end{tabular}

\vspace{2pt}{\footnotesize On ${512\times256\times256}$ elements the final evaluation took 0.8 to 2.3 minutes of each total. Totals exceed 300 times the median because the time per design iteration varies during a run and the totals include setup and the final evaluation. The L-bracket counts the elements of its enclosing box; 3.35 million of them are active. $^{1}$Regularization of the MMA scaled (Amendment 11); the two rows above used the value of the released code and formed no mechanism}\par
\end{table}

On the fine mesh each optimization took 9 to 35 minutes, with at most 1.64\,GiB of GPU memory
(Table~\ref{tab:time}). The cantilever on the ${512\times256\times256}$ mesh has 33.5 million elements and 101.6 million degrees of freedom. It took a median of 17.0\,s per design iteration and 86.5 minutes for 300 design iterations and the final evaluation,
with a peak of 9.43\,GiB of GPU memory. The five additional cases on this mesh met the execution criteria of the protocol and took 78 to 223 minutes, with at most 9.93\,GiB. The robust cantilever and the two infill cases led to the same conclusions as on the fine mesh, although the robust design has a
different arrangement of members (Online Resource~1, Fig.~S2). The eroded robust cantilever had 0.728 times the compliance of the eroded single projection cantilever (0.692 on the fine mesh), and a morphological opening removed less solid from it (0.24 against 0.26\,\%; 22.5 against 24.2\,\% with the radius of 11 elements of Amendment 10). The infill cantilever had 0.897 times the lateral compliance of its reference, at 1.171 times its nominal compliance (0.922 and 1.158 on the fine mesh).

With the regularization of the released code, neither inverter formed a mechanism on this mesh within 300 design iterations: the output
displacement fell below 0.2\,\% of its initial magnitude, with the wrong sign in the robust design, and the volume
fraction ended at 0.06 and 0.08 against a bound of 0.25. On the ${256\times128\times128}$ mesh the same output grew to
about 20 times its initial magnitude. A test on the coarse mesh points to the optimizer (Amendment 10; Online Resource~1, Table~S4). The regularization of the MMA is fixed, whereas the element sensitivities fall as the number of elements grows. The inverter on the ${256\times128\times128}$ mesh had also lost its load path, from design iteration 2 to 80, before it formed a mechanism; on the largest mesh it did not recover. To give the regularization on the coarse mesh the relative weight it has on the finer meshes, the test multiplied it by the cube of the mesh ratio (18.96 and 151.70). The design then lost its load path for 51 and 196 design iterations instead of one, with a volume history close to that of the finer meshes. Amendment 11 then divided the regularization on the largest mesh by $(512/96)^3 = 151.7$, without changing the released code and with criteria fixed before the runs. Both inverters met these criteria (Online Resource~1, Fig.~S2 and Table~S4). Each lost its load path for one design iteration, as on the coarse mesh, and the volume fraction ended at its bound of 0.25. The output reached 26.6 and 25.2 times its initial magnitude. Eroding to $\eta = 0.70$ removed 99.6\,\% of the output of the single projection inverter and 2.0\,\% of the output of the robust one (98 and 1.7\,\% on the fine mesh). These runs took 125 and 309 minutes. Only this value of the regularization was tried, and the released code keeps $10^{-5}$.

\begin{figure*}[tp]
\centering
\includegraphics[width=0.96\textwidth]{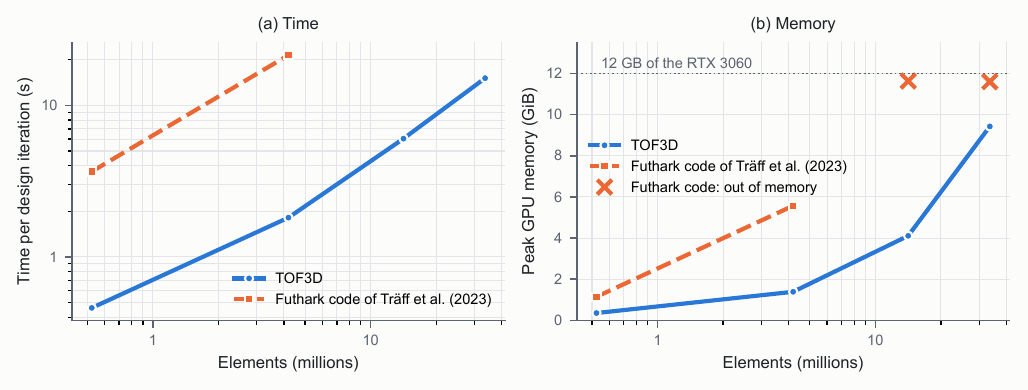}
\caption{TOF3D and the open Futhark code of \citet{traff2023gpu} on the same RTX 3060 and the same cantilever, from
${128\times64\times64}$ to ${512\times256\times256}$ elements. (a) Median time per design iteration over iterations 2 to
20 and three repetitions; (b) peak GPU memory. Crosses: the Futhark code stopped when the memory of the card ran
out}\label{fig:gpucodes}
\end{figure*}

On the same card and problem, the Futhark code of \citet{traff2023gpu} reproduced the compliance of the first
design iteration of the double precision reference to within ${4.4\times10^{-5}}$, and TOF3D to within
${1.7\times10^{-7}}$. With 0.5 and 4.2 million elements, TOF3D took 0.46 and 1.8\,s per design iteration, and the
Futhark code 3.7 and 21\,s (Fig.~\ref{fig:gpucodes}). The three repetitions of each code differed by less than 2\,\%. Per linear solution, the Futhark code took a median
of 16.5 and 15 conjugate gradient iterations on these two meshes, and TOF3D 13.5 and 10.
With 14.2 and 33.5 million elements, the Futhark code ran out of the 12\,GB of memory, whereas TOF3D took 6.0 and
15\,s per design iteration with at most 9.4\,GiB. The protocol cantilever of Table~\ref{tab:time}, with a larger
filter radius and a different load, took 17.0\,s on the same mesh. The Futhark code applies the operator on the fine grid in double precision,
which the RTX 3060 runs at a small fraction of its single precision rate, so on a data center GPU such as the A100 the ranking of the two codes may
differ. TOF3D reaches single precision speed and memory without the error of Table~\ref{tab:precision} because of the
difference form of Eq.~(\ref{eq:diff}).

\begin{table*}[tp]
\caption{Recent codes for large three-dimensional topology optimization, as reported in each study}\label{tab:codes}
\small\begin{tabular}{@{}P{0.2\textwidth}P{0.2\textwidth}P{0.33\textwidth}P{0.19\textwidth}@{}}
\toprule
Study & Hardware & Problems & Largest model \\
\midrule
\citet{traff2023gpu} & A100 GPU, 80\,GB & Minimum compliance & 65.5\,M elements \\
\citet{zhang2023homogenization} & GTX 1080 Ti GPU, 11\,GB & Inverse homogenization & 134.2\,M elements \\
\citet{zhao2024fiber} & V100 GPU, 32\,GB & Fiber-reinforced composites & 33.6\,M elements \\
\citet{wang2025matrixfree} & Six-core CPU, 64\,GB & Minimum compliance, porous infill & 128\,M elements \\
\citet{jia2026largescale} & RTX 4090 GPU, 24\,GB & Length scale and local volume constraints & 151\,M degrees of freedom \\
This work (TOF3D) & RTX 3060 GPU, 12\,GB & Stress constraints, robustness, compliant mechanisms, porous infill & 33.5\,M elements (cantilever, infill, inverter) \\
\botrule
\end{tabular}
\end{table*}

Table~\ref{tab:codes} places TOF3D among recent codes for large three-dimensional problems. On data center GPUs, \citet{traff2023gpu}
optimized a cantilever with 65.5 million elements and a filter radius of 2.5 elements on an 80\,GB A100, in
7\,380\,s for 100 design iterations, and \citet{zhao2024fiber} reached 33.6 million elements on a 32\,GB V100 at about
70\,s per design iteration. On a consumer card, \citet{zhang2023homogenization} ran inverse homogenization with 134.2 million elements on
an 11\,GB GTX 1080 Ti in less than 40\,s per iteration. Without a GPU, \citet{wang2025matrixfree} optimized a bracket with 11 million
elements and four load cases in Matlab on a six-core CPU, in 7.6\,h for 50 design iterations, with a residual
tolerance of $10^{-3}$; because this code runs on the CPU, it was not timed on the RTX 3060. The filters, solver tolerances, floating point precision and problems differ among these
studies, so the numbers do not rank the codes. TOF3D combines stress constraints, robustness,
mechanisms and infill in one code, in single precision on a 12\,GB consumer GPU, with its responses cross-checked
against a commercial code.

\subsection{Implications for learned surrogates}\label{sec:rsurrogates}

The cost of conventional optimization is the reference against which learned surrogates of topology optimization are judged.
\citet{sigmund2022benchmarking} asks learning-based studies to report a fair break-even threshold, the number of
optimizations needed to train a network weighed against its generality, and refers to the review of
\citet{woldseth2022use} for the issues involved. On the fine mesh, TOF3D produces a design that meets its constraints after 300 design iterations,
with 4.2 to 5.2 million elements, in 9 to 35 minutes on a 12\,GB consumer GPU. Its single precision
responses agree with double precision to within ${1.3\times10^{-5}}$, and its double precision solutions agree with a
commercial code to within ${4.1\times10^{-10}}$. A surrogate trained on such designs is justified only if its
predictions approach this accuracy for boundary conditions, domains and volume fractions outside its training set. It must also recover the cost of generating its training designs with a solver like this one. At these times and this
precision the break-even threshold is likely to be high, although no surrogate was tested here. The present results are consistent with the view of
\citet{sigmund2022benchmarking} and \citet{woldseth2022use} that efficient conventional optimization on accessible
hardware is the reference for learning-based methods.

\section{Conclusion}\label{sec:concl}

This work proposed TOF3D, an open source Python code for density-based topology optimization in three dimensions, in
single precision, on one consumer GPU. The code covers four classes of problems: stress constraints, robustness against manufacturing
errors, compliant mechanisms and porous infill. The main conclusions are as follows.
\begin{itemize}
\item The four classes ran on a 12\,GB RTX 3060. With 4.2 to 5.2 million elements, each optimization took 9 to 35
minutes and at most 1.64\,GiB of GPU memory; with 33.5 million elements, the optimizations took 78 to 309 minutes and at
most 9.9\,GiB.
\item On the same RTX 3060 and problem, TOF3D took 0.46 and 1.8\,s per design iteration with 0.5 and 4.2 million
elements. An open Futhark code that applies its operator in double precision took 3.7 and 21\,s. That code ran
out of memory above 4.2 million elements; TOF3D reached 33.5 million with 9.4\,GiB.
\item The multigrid preconditioner kept the conjugate gradient method at 12 to 18 iterations from 0.2 to 101.6
million degrees of freedom, and solved 8 to 12 times faster than block Jacobi on the larger meshes.
\item With the same element, TOF3D agreed with Ansys Mechanical APDL to within ${4.1\times10^{-10}}$ in displacements on
ten designs. The fully integrated element of TOF3D differs from SOLID185 with B-bar integration by 0.32 to 1.0\,\% in
compliance.
\item Rounding the element matrix to single precision caused errors of up to 5.4\,\% that grew with the mesh.
Forming the element products from displacement differences reduced them to ${1.3\times10^{-5}}$ at no cost in time.
\item Of the four hypotheses fixed before the runs, all four were confirmed on the fine mesh and two on the coarse
mesh. On the coarse mesh the compliance increase caused by the stress constraint depended on the optimization path, and
the single projection inverter failed only beyond the erosion threshold of the protocol. With 33.5 million elements
the robust and infill cantilevers led to the same conclusions as on the fine mesh. The inverters formed a mechanism on this mesh only after the fixed regularization of the MMA was scaled with the number of elements.
\end{itemize}
The code, the protocol and the results are deposited as a reference for other GPU codes and for learned surrogates.

\backmatter
\bmhead{Supplementary information}
Online Resource~1 contains sections of the infill design (Fig.~S1), the dated amendments to the protocol
(Table~S1), the secondary analysis of the coarse L-bracket (Table~S2) and the designs on the ${512\times256\times256}$ mesh
(Fig.~S2), and the analyses of Amendments 10 and 11 (Tables~S3 and~S4).

\bmhead{Acknowledgments}

This work was developed within the PROPEMM graduate program at IFES. The authors thank
the Laborat\'orio de Simula\c{c}\~ao Computacional, IFES Campus Aracruz, for computing
infrastructure.

\bmhead{Statements and declarations}

\textbf{Data and code availability.}
The code, under the MIT license, the protocol with its dated amendments, the scripts of every check and figure and
the final designs are deposited on Zenodo (\href{https://doi.org/10.5281/zenodo.24050327}{https://doi.org/10.5281/zenodo.24050327}). Ansys Mechanical APDL is commercial software:
the input files written for the cross-check and the results extracted from it are deposited, but not its binary
result files.

\textbf{Conflict of interest.} The authors declare no competing interests.

\textbf{Use of language models.} The use of large language models is declared in Sect.~\ref{sec:env}.

\textbf{Funding.} The article processing charge is expected to be covered by the Coordena\c{c}\~ao de Aperfei\c{c}oamento de Pessoal de N\'ivel Superior~-- Brasil (CAPES) (ROR identifier: 00x0ma614) under the CAPES--Springer Nature agreement, subject to CAPES authorization after acceptance. The authors received no other specific funding for this work.

\textbf{Author contributions.} Conceptualization: L.T.P., R.M.F.; Methodology, Software, Investigation, Data curation,
Formal analysis, Visualization, Writing -- original draft, Writing -- review \& editing: L.T.P.; Supervision, Validation: R.G.M.A., R.M.F.

\textbf{Ethics approval, consent to participate and consent for publication.}
Not applicable. The study uses only simulated structural data.

\bmhead{Replication of results}
The deposit contains the code, the protocol with its dated amendments, the scripts of every check and figure, and
the final designs. Its README lists the required packages and maps each table and figure of this article to its data and script. After the packages are installed, the command \texttt{python tof3d\_examples.py C\_std -{}-size 96 -{}-iters 20 -{}-smoke} runs a short test optimization (option \texttt{-{}-smoke}) of the single projection cantilever (case \texttt{C\_std}) on a ${96\times48\times48}$ grid, a size at which a new problem can be tried first. The command \texttt{bash tests/run\_all.sh python} repeats the checks of Sect.~\ref{sec:rverif}. The gradient checks and the double precision cross-check with MAPDL repeat to round-off on any
computer. The optimized designs repeat on the same GPU and software versions; on other hardware, floating point
reductions in a different order can change the optimization path and, slightly, the final design. Times and memory
depend on the GPU.

\bibliography{references}

\end{document}